\pdfoutput=1
\documentclass[11pt,english]{article}
\usepackage[T1]{fontenc}
\usepackage[latin9]{inputenc}
\usepackage{array}
\usepackage{verbatim}
\usepackage{multirow}
\usepackage{amsmath}
\usepackage{amssymb}
\usepackage{graphicx}
\usepackage{esint}

\makeatletter

\providecommand{\tabularnewline}{\\}

\@ifundefined{date}{}{\date{}}
\date{}
\usepackage{cite}

\makeatother

\usepackage{babel}
\begin{document}

\title{Alchemical Response Parameters from an Analytical Model of Molecular
Binding}

\author{Emilio Gallicchio%
\thanks{Department of Chemistry, Brooklyn College, 2900 Bedford Avenue, Brooklyn,
New York 11210, and Ph.D. Program in Biochemistry, The Graduate Center
of the City University of New York, New York, NY 10016, and Ph.D.
Program in Chemistry, The Graduate Center of the City University of
New York, New York, NY 10016. E-mail: egallicchio@brooklyn.cuny.edu%
}}
\maketitle
\begin{abstract}
We present a parameterized analytical model of alchemical molecular
binding. The model describes accurately the free energy profiles of
linear single-decoupling alchemical binding free energy calculations.
The parameters of the model, which are physically motivated, are obtained
by fitting model predictions to numerical simulations. The validity
of the model has been assessed on a set of host-guest complexes. The
model faithfully reproduces both the binding free energy profiles
and the probability densities of the perturbation energy as a function
of the alchemical progress parameter $\lambda$. The model offers
a rationalization for the characteristic shape of the free energy
profiles. The parameters obtained from the model are potentially useful
descriptors of the association equilibrium of molecular complexes.
\end{abstract}

\section{Introduction}

The primary goal of a quantitative model of molecular binding is to
provide an estimate of the standard free energy of binding, $\Delta G_{b}^{\circ}$,
or, equivalently, of the equilibrium constant, $K_{b}$, for the association
equilibrium $R+L\leftrightharpoons RL$, between two molecules $R$
and $L$. For example, the binding of a drug molecule to a receptor.
A brute-force molecular simulation approach to the calculation of
the binding constant, based on following the motion of the ligand
in and out of the receptor, is generally not feasible due to the long
times between binding and unbinding events.\cite{pan2013molecular}
To overcome obstacles such as this, biased methods have been developed
to accelerate the dynamics of association and obtain the free energy
profile of ligand binding along pathways in and out of the receptor.\cite{gumbart2013efficient,limongelli2013funnel,cavalli2014investigating,comer2014adaptive,di2014mechanistic,tiwary2015kinetics,sandberg2015efficient,miao2015gaussian,saglam2016highly}

Alchemical descriptions of the binding equilibrium provide an alternative
to the study of physical binding/unbinding paths.\cite{Chodera:Mobley:cosb2011,Gallicchio2011adv,mobley2012perspective,cuendet2012alchemical}
The idea is that, because a free energy change depends only on the
end states, the bound and unbound states of the molecular system can
be connected by any thermodynamic path, whether physical or unphysical.
In alchemical methods, the potential energy function is modified parametrically
in a series of steps traced by a progress parameter $\lambda$ to
go from a description of the unbound state to that of the bound state.
These methods effectively ``grow'' the ligand in place within the
binding site. The field has a long history, \cite{McCammon:84,Jorgensen1989,Payne:Matubayasi:Reed:Levy:97,Gallicchio:Kubo:Levy:98}
but only relatively recently it has converged into an unified statistical
thermodynamics theory of bimolecular binding. \cite{Gilson:Given:Bush:McCammon:97,Boresch:Karplus:2003,Chipot:Pohorille:book:2007,Gallicchio2011adv}
The double-decoupling method,\cite{Gilson:Given:Bush:McCammon:97,Chodera:Mobley:cosb2011,deng2011elucidating}
which is used to compute absolute binding free energies, is so called
because it involves free energy calculations to decouple the ligand
to an intermediate gas phase from the bound and solution states of
the ligand. Free energy perturbation methods,\cite{lybrand:mccammon1986,Michel2007,oostenbrink2011free,wang2015accurate}
are suitable for the analysis of relative binding, such as in ligand
optimization.

We have developed an alchemical single-decoupling methodology, based
on an implicit description of the solvent,\cite{Gallicchio2009} that
enables the transfer of the ligand directly into the binding site
rather than through multiple thermodynamic pathways.\cite{Gallicchio2010,Gallicchio2012a,Gallicchio2012b}
Among other advantages, the single-decoupling approach leads naturally
to a statistical representation of the equilibrium in terms of probability
distributions of the binding energy. For example, it is possible to
relate the binding free energy to the probability distribution, $p_{0}(u)$,
of the the binding energy in the absence of receptor-ligand interactions.\cite{Gallicchio2011adv}

Analogously to approaches based on physical binding pathways, alchemical
binding free energy calculations yield free energy profiles along
the thermodynamic transformations. Alchemical free energy profiles
are functions of the alchemical charging parameter $\lambda$, rather
than, for instance, the ligand-receptor distance. A typical alchemical
calculation involves collecting distributions of perturbation energies
as a function of the alchemical parameter $\lambda$. These are merged
together by thermodynamic reweighting algorithms\cite{Shirts2008a,Tan2012}
to yield the free energy profile along $\lambda$. Typically, only
the difference between the end points of the free energy profile,
which is the binding free energy, is considered. However the shape
of the free energy profile can also yield useful information regarding
the physical characteristics of the molecular complex. For example,
a quadratic dependence on $\lambda$, typical of linear response,
is often observed during the alchemical transformation. 

In this work we present a method to relate the shape of the free energy
profile to physical observables of the complex. Working within the
single-decoupling framework, we develop an analytic probabilistic
model of binding and construct a procedure to estimate the parameters
of the model from the results of alchemical molecular simulations.
The model is based on the the statistics of ligand-receptor interaction
energies when the ligand uniformly explores the binding site volume
as if the receptor atoms were not present. This general strategy has
a long history in the treatment of solvation (examples are scaled
particle theory, particle insertion, and information/fluctuation theories
\cite{Widom1982,Pratt1992,Hummer:Pratt:96,Simonson:99,Huang:Chandler:2002})
but it has not been fully explored to study molecular recognition.
The main distinction is that a receptor, unlike a homogeneous solvent,
has a specific shape and distribution of interaction sites. We show
that the single decoupling theory offers a useful starting point to
think about this problem.

\section{Theory and Methods}

\subsection{Statistical mechanics theory of non-covalent molecular association}

The standard free energy of binding, $\Delta G_{b}^{\circ}$, between
a receptor $R$ and a ligand $L$ is given by
\begin{equation}
\beta\Delta G_{b}^{\circ}=-\ln K_{b},\label{eq:DG0}
\end{equation}
where $\beta=1/(k_{B}T)$, $T$ is the absolute temperature, $k_{B}$
is Boltzmann's constant and $K_{b}$ is the dimensionless binding
constant that, assuming ideal solutions, is expressed as 
\begin{equation}
K_{b}=\frac{[RL]/C^{\circ}}{([R]/C^{\circ})([L]/C^{\circ})},\label{eq:Kb-def}
\end{equation}
where $[\ldots]$ are equilibrium concentrations and $C^{\circ}$
is the standard state concentration (conventionally set as 1M or 1
molecule/1668 \AA$^{3}$). 

In a widely employed classical statistical mechanics theory of non-covalent
association,\cite{Gilson:Given:Bush:McCammon:97,Gallicchio2011adv}
the binding constant is expressed as 
\begin{equation}
K_{b}=C^{\circ}V_{{\rm site}}\langle e^{-\beta\Delta U}\rangle_{0},\label{eq:Ka_impl}
\end{equation}
where $U(x,\zeta)=V(x,\zeta)+W(x,\zeta)$ is the effective potential
energy function of the receptor-ligand complex expressed in terms
of the internal degrees of freedom, $x$, of receptor and ligand,
and the external degrees of freedom (i.e.~overall translation and
rotations),\cite{Boresch:Karplus:2003} $\zeta$, of the ligand with
respect to the receptor, and $\Delta U(x,\zeta)=U(x,\zeta)-U_{0}(x)$
is the binding energy of the complex in conformation $(x,\zeta)$,
where $U_{0}(x)$ is the effective potential energy of the system
when receptor and ligand are at infinite separation. $V_{{\rm site}}$
is the chosen volume of the binding site, that is the volume of the
region of positions and orientations $\zeta$ of the ligand relative
to the receptor which are considered to correspond to the bound state
of the complex.%
\footnote{Eq.~(\ref{eq:Ka_impl}) refers to the case in which only overall
translations are used to define the binding site volume. In general,
a term corresponding to the integration over orientational degrees
of freedom is also present.\cite{Boresch:Karplus:2003,Gallicchio2011adv} %
} The average $\langle\ldots\rangle_{0}$ in Eq.~(\ref{eq:Ka_impl})
is conducted over the decoupled equilibrium ensemble corresponding
to $U_{0}(x)$, in which receptor and ligand do not interact, while
the ligand samples uniformly the binding site volume. Here $V(x,\zeta)$
is the potential energy of the system and $W(x,\zeta)$ is the solvent
potential of mean force, which represents the solvation free energy
of the complex in conformation $(x,\zeta)$.\cite{Gallicchio2011adv} 

Inserting Eq.~(\ref{eq:Ka_impl}) into Eq.~(\ref{eq:DG0}) yields
\begin{equation}
\beta\Delta G_{b}^{\circ}=-\ln C^{\circ}V_{{\rm site}}+\beta\Delta G_{{\rm exc.}},\label{eq:DG02}
\end{equation}
where $-k_{B}T\ln C^{\circ}V_{{\rm site}}$ is the concentration-dependent
component of the standard free energy of binding independent of the
specific form of the potential energy, and
\begin{equation}
\beta\Delta G_{{\rm exc.}}=-\ln\langle e^{-\beta\Delta U}\rangle_{0}\label{eq:DGexc}
\end{equation}
is the excess free energy of the complex. In the following we will
focus on the excess component of the standard free energy of binding
and, to simplify the notation, we will denote the excess free energy
as simply $\Delta G$ and we will measure all energies and free energies
in units $k_{B}T$ thereby eliminating factors of $\beta$ throughout.

\subsection{Alchemical binding free energy methods}

Molecular simulations aimed at computing the excess free energy of
binding by means of Eqs.~(\ref{eq:DG02}) and (\ref{eq:DGexc}) and
are referred to as ``alchemical'' in that they sample the unphysical
uncoupled state in which receptor and ligand, while being close to
each other, they behave as if the other were not present. In practice,
Eq.~(\ref{eq:DGexc}) converges very slowly because, due to atomic
overlaps, in the uncoupled state large and positive values of $\Delta U$
(and, consequently, negligibly small values of $\exp(-\Delta U)$)
are much more likely to be sampled than favorable ones, causing the
average to be dominated by the infrequent occurrences of overlap-free
configurations. To overcome this obstacle, it is common to adopt a
stratification scheme based on an alchemical hybrid potential $U(x,\zeta;\lambda)$,
dependent on an alchemical progress parameter $\lambda$ conventionally
ranging from 0 and 1, which implies a $\lambda$-dependent excess
free energy defined as 
\begin{equation}
\Delta G(\lambda)=-\ln K(\lambda),\label{eq:DGexc-1}
\end{equation}
where
\begin{equation}
K(\lambda)=\langle e^{-\Delta U(\lambda)}\rangle_{0}\ ,\label{eq:Klambdadef}
\end{equation}
is the $\lambda$-dependent binding constant, and where, using the
notation introduced above, $\Delta U(\lambda)=U(x,\zeta;\lambda)-U_{0}(x,\zeta)$
is the perturbation energy at $\lambda$ for the complex in conformation
$(x,\zeta)$. In the following we will refer to $\Delta G(\lambda)$
as the \textit{alchemical free energy profile} and $K(\lambda)$ as
the \textit{binding constant profile}.

The stratification approach above leads to the familiar computational
algorithms for the calculation of free energy differences based on
the accumulation of the effects of small progressive increments of
$\lambda$. For instance, Eq.~(\ref{eq:Klambdadef}) is easily generalized
to yield an expression of the ratio of equilibrium constants at nearby
values of $\lambda$:
\begin{equation}
\frac{K(\lambda')}{K(\lambda)}=\langle e^{-[\Delta U(\lambda')-\Delta U(\lambda)]}\rangle_{\lambda}\ ,\label{eq:Klambda_ratio}
\end{equation}
which is the basis of the Free Energy Perturbation (FEP) method. It
should be noted that, while Eq.~(\ref{eq:Klambda_ratio}) is mathematically
exact, modern numerical implementations of FEP employ more efficient
BAR and MBAR free energy estimators.\cite{Shirts2008a} Similarly,
inserting Eq.~(\ref{eq:Klambdadef}) into Eq.~(\ref{eq:DGexc-1})
and differentiating with respect to $\lambda$, leads to the well-known
Thermodynamic Integration (TI) formula:
\begin{equation}
\frac{d\Delta G(\lambda)}{d\lambda}=\langle\frac{\partial U(\lambda)}{\partial\lambda}\rangle_{\lambda}\label{eq:TI}
\end{equation}
which, when integrated, yields the free energy profile.

Being related to ensemble averages, it is helpful for the current
purpose to note that both the FEP and TI formulas can be expressed
in terms of probability density functions. For instance, Eq.~(\ref{eq:Klambdadef})
can be rewritten as
\begin{equation}
K(\lambda)=\int_{-\infty}^{+\infty}d(\Delta U_{\lambda})e^{-\Delta U_{\lambda}}p_{0}(\Delta U_{\lambda})\ ,\label{eq:Klambdadef-prob}
\end{equation}
where $p_{0}(\Delta U_{\lambda})$ is the probability density of the
perturbation energy, $\Delta U(\lambda)$, at $\lambda$ in the $\lambda=0$
ensemble. Analogously, denoting $u(\lambda)=\partial U(\lambda)/\partial\lambda$,
Eq.~(\ref{eq:TI}) is rewritten as
\begin{equation}
\frac{d\Delta G(\lambda)}{d\lambda}=\int_{-\infty}^{+\infty}du\, up_{\lambda}(u)\label{eq:TI-prob}
\end{equation}
where $p_{\lambda}(u)$ is the probability density of the $\partial U/\partial\lambda$
function in the ensemble at $\lambda$.

\subsection{Linear alchemical transformations}

Eqs.~(\ref{eq:Klambdadef-prob}) and (\ref{eq:TI-prob}) take a particular
convenient form when the alchemical potential energy function $U(x,\zeta;\lambda)$
is linear with respect to $\lambda$:
\begin{equation}
U(x,\zeta;\lambda)=U_{0}(x)+\lambda u(x,\zeta)\label{eq:U-lambda-linear}
\end{equation}
where $U_{0}(x)$ is the potential energy of the decoupled state and
$u(x,\zeta)$ is the so-called binding energy function of the complex,
which, critically, is assumed here independent of $\lambda$. It is
straightforward to show that for an alchemical potential of the form
(\ref{eq:U-lambda-linear}) the perturbation potential is proportional
to the binding energy function 
\begin{equation}
\Delta U(x,\zeta;\lambda)=\lambda u(x,\zeta)\label{eq:lambda-u}
\end{equation}
and that the $\lambda$-derivative employed in the TI formula is independent
of $\lambda$ and is given by the binding energy function itself:
\begin{equation}
\frac{\partial U(x,\zeta;\lambda)}{\partial\lambda}=u(x,\zeta)\:.\label{eq:dudl=00003Du}
\end{equation}

Inserting Eq.~(\ref{eq:lambda-u}) into Eq.~(\ref{eq:Klambdadef-prob})
we obtain 
\begin{equation}
K(\lambda)=\int_{-\infty}^{+\infty}du\: e^{-\lambda u}p_{0}(u)\ ,\label{eq:Klambdadef-prob-linear}
\end{equation}
where $p_{0}(u)$, which plays a central role in this work, is the
probability density of the binding energy function in the uncoupled
state, that is in the state in which the ligand is uniformly distributed
in the binding site region and receptor and ligand do not interact
with each other. 

Mathematically, Eq.~(\ref{eq:Klambdadef-prob-linear}) expresses
the fact that the binding constant profile $K(\lambda)$ is given
by the two-sided Laplace transform of $p_{0}(u)$. In turn, the binding
free energy profile $\Delta G(\lambda)$ is related to the $K(\lambda)$
by Eq.~(\ref{eq:DGexc-1}), and the excess binding free energy is
$\Delta G(\lambda=1)$. Finally, the Potential Distribution Theorem\cite{PDTbook:2006}
provides a relationship between $p_{0}(u)$ and the binding energy
distributions at any other value of $\lambda$:
\begin{equation}
p_{\lambda}(u)=e^{\Delta G(\lambda)}e^{-\lambda u}p_{0}(u).\label{eq:plu}
\end{equation}
It is therefore apparent that knowledge of $p_{0}(u)$ determines
all of the other quantities that characterize the alchemical transformation,
including the binding free energy profile and the binding free energy.
In this respect the function $p_{0}(u)$ serves the same role in the
alchemical theory of binding that the density of states $\Omega(E)$
plays in classical statistical mechanics. For instance note the parallel
between Eq.~(\ref{eq:plu}) and the well know Boltzmann's relationship
$p_{\beta}(E)\propto\exp[-\beta E]\Omega(E)$, which gives the energy
distribution of a system at any temperature given the density of states. 

The main aim of the work presented here is to develop a probabilistic
analytical model for $p_{0}(u)$ from which to derive all of the other
quantities discussed above and, conversely, to estimate the parameters
of the model against the results of alchemical molecular simulations.

\subsection{Statistical model for $p_{0}(u)$}

\begin{figure}
\begin{centering}
\includegraphics[scale=0.5]{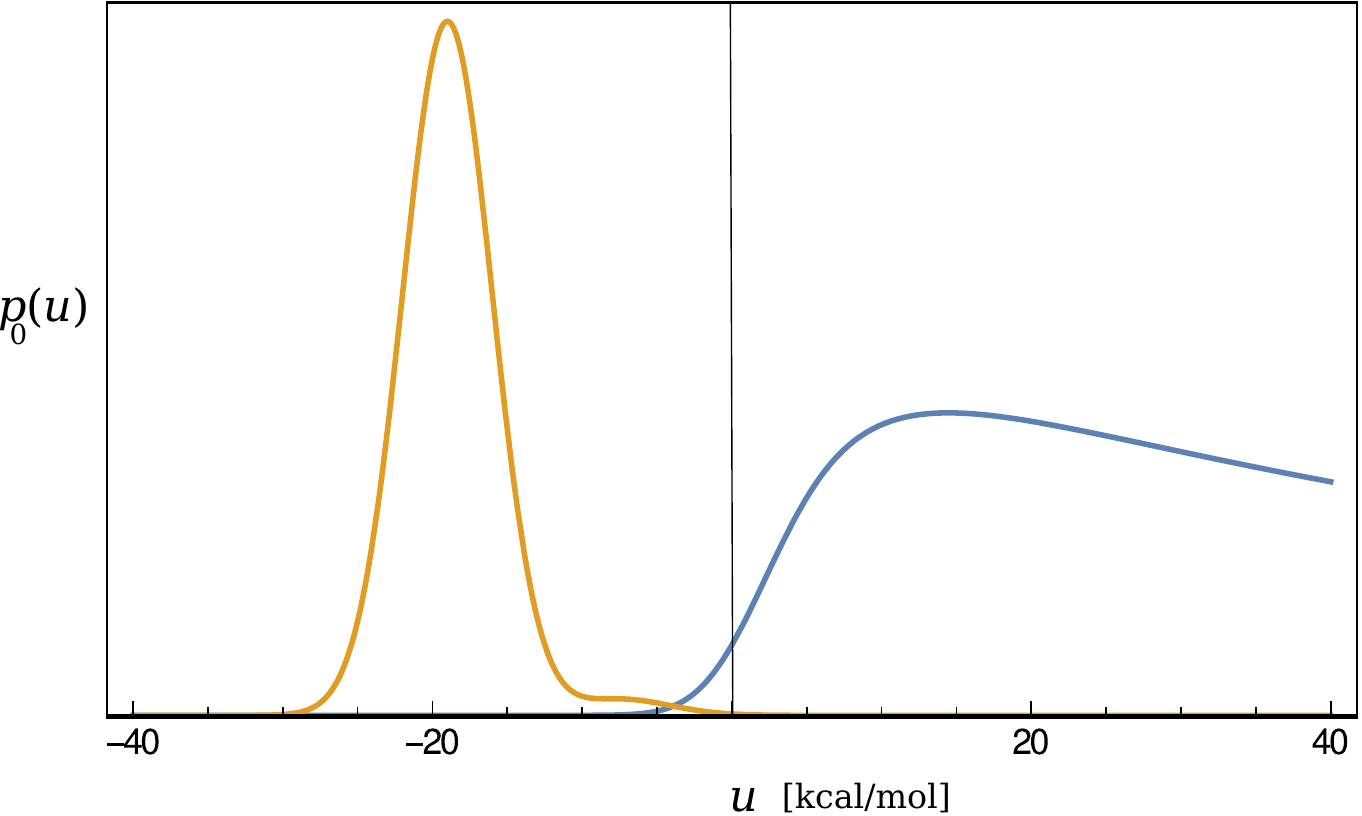}
\par\end{centering}

\caption{\label{fig:p0u-1}$p_{0}(u)$ (blue, curve on the left) and $p_{1}(u)$
(yellow, curve on the right) from Eq.~(\ref{eq:p0(u)conv2}) for
$\bar{u}_{B}=-10$, $\sigma_{B}=3$, $\epsilon_{LJ}=1$, $\tilde{u}_{c}=10$,
$n_{l}=2$ and $\tilde{p}_{{\rm core}}=10^{-6}$. The scale of the
$y$-axis is arbitrary and probability densities are unnormalized.
Energy values are expressed in units of $k_{B}T$.}
\end{figure}

In this section we turn to derivation of a model for the probability
distribution, $p_{0}(u)$, of the binding energy in the uncoupled
ensemble at $\lambda=0$, that is in the state when the ligand and
the receptor are not interacting. Note the key distinction between
the state from which samples are collected (the uncoupled ensemble),
and the quantity being sampled (the binding energy function): we are
interested in the distribution of binding energies, which are in general
not zero, when receptor and ligand configurations are sampled in the
absence of receptor-ligand interactions. As illustrated in Fig.~\ref{fig:p0u-1},
due to the fact that in absence of interactions clashes between ligand
and receptor atoms are likely, $p_{0}(u)$ is characterized by a long
tail at large and positive values of the binding energy. $p_{0}(u)$
has also a much smaller, but finite, tail at favorable binding energies.
The low energy tail of $p_{0}(u)$ is amplified by the $\exp(-u)$
exponential term, to yield, through Eq.~(\ref{eq:plu}), the expected
distribution of binding energies in the bound state narrowly centered
around a favorable mean binding energy (see Fig.~\ref{fig:p0u-1}).

\begin{figure}
\begin{centering}
\includegraphics[scale=0.5]{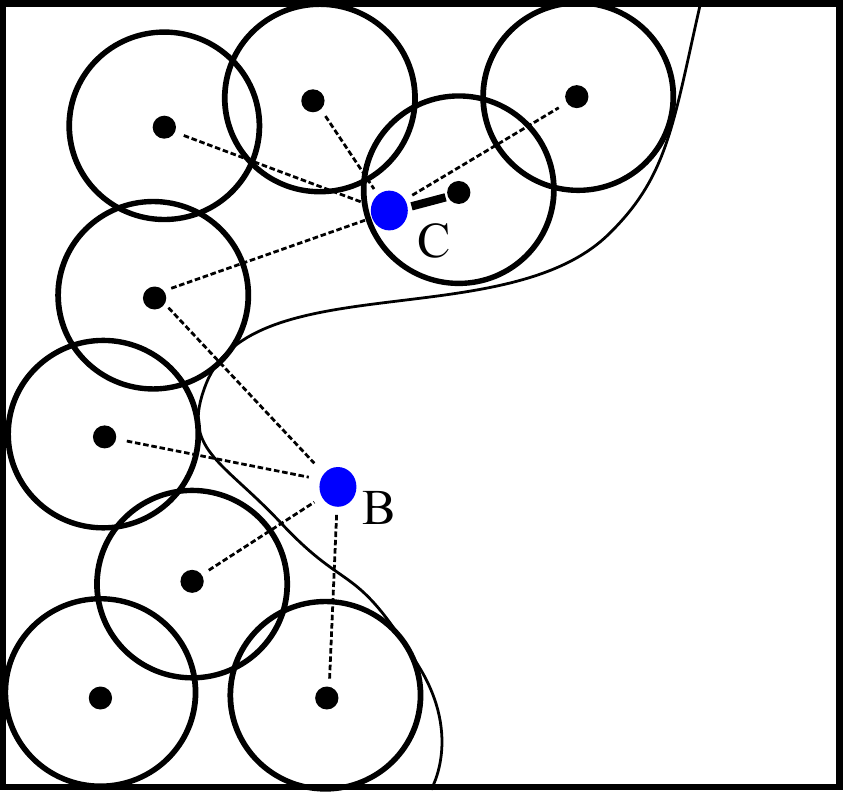}
\par\end{centering}

\caption{An illustration of the ligand (blue) interacting with the atoms of
a receptor. In location B, outside the core region of the receptor
site, the interaction energy between ligand and receptor is the sum
of many, long ranged, pair interactions. In location C within the
core of a receptor atom, instead, one repulsive interaction dominates
all of the others.\label{fig:Receptor}}
\end{figure}

To start thinking about a functional form for $p_{0}(u)$, here we
consider a monoatomic ligand. The model will be generalized to arbitrary
ligand molecules later in this section. Consider Fig.~\ref{fig:Receptor},
which depicts the binding site volume containing receptor atoms arranged
in some configuration, with a monoatomic ligand placed in two alternative
positions (blue spheres, labeled ``B'' and ``C''). The binding
site volume here is a represented as a rectangular box, although the
following arguments apply to any definition of the binding site volume.
Because at $\lambda=0$ it does not interact with receptor atoms,
the ligand occupies the binding site with uniform probability. The
effective interaction energy between the ligand and the receptor is
the sum of many individual interatomic interactions. In regions of
the binding site sufficiently removed from the interior of the atoms
of the receptor, as for example location ``B'' in Fig.~\ref{fig:Receptor},
the interaction energy is approximately the result of many, relatively
weak and favorable pair-wise interactions of similar magnitude. This
mode of interaction describes the behavior of $p_{0}(u)$ at favorable
values of the binding energy.

When, instead, the ligand is found within the inner core of a receptor
atom, as location ``C'' in Fig.~\ref{fig:Receptor}, the repulsion
energy of that individual interaction dominates all of the others.
This interaction mode, expected to important to describe the high
energy tail of $p_{0}(u)$. The atomic core of an atom is considered
here as its most immediate region where its interaction potential
dominates over other longer-ranged interactions. Because receptor
atoms can not interpenetrate each other to more than a certain degree,
strong repulsive interactions can be understood as the result of a
single pair interaction rather than of cooperative contributions of
many interactions.

To model these two distinct statistical behaviors, it is useful to
think of the ligand-receptor binding energy as the results of two
contributions
\begin{equation}
u=u_{C}+u_{B},
\end{equation}
where $u_{C}$ represent the \textit{collisional} component, which
corresponds to short-ranged repulsive interactions which dominate
within the atomic cores and are represented by a single pair interaction
at a time, and $u_{B}$ is the \textit{background} component given
by the sum of contributions of many weak and favorable longer-ranged
pair interactions. Motivated by the central limit theorem, we model
the probability distribution of the background component by a Gaussian
distribution:
\begin{equation}
p_{B}(u_{B})=g(u_{B};\bar{u}_{B},\sigma_{B})=\frac{1}{\sqrt{2\pi\sigma_{B}^{2}}}\exp\left[-\frac{(u_{B}-\bar{u}_{B})^{2}}{2\sigma_{B}^{2}}\right],\label{eq:PB(u)}
\end{equation}
where $\bar{u}_{B}$ is the mean and $\sigma_{B}$ is the standard
deviation of the distribution.

The statistics of the collisional energy $u_{C}$ in the region outside
the atomic cores is unimportant since $u_{C}$ is much smaller than
$u_{B}$ there. On the other hand, when inside one of the atomic cores,
$u_{C}$ is large and positive. Here, for a single pair collision,
we assume a repulsive pair potential of the Lennard-Jones (LJ), Weeks-Chandler-Andersen
(WCA) form
\begin{equation}
u_{WCA}(r)=\begin{cases}
4\epsilon_{LJ}\left[\left(\frac{\sigma_{LJ}}{r}\right)^{12}-\left(\frac{\sigma_{LJ}}{r}\right)^{6}\right]+\epsilon, & r<2^{1/6}\sigma_{LJ}\\
0 & r>2^{1/6}\sigma_{LJ}\:,
\end{cases}\label{eq:WCApot}
\end{equation}
which, as shown in the Appendix, within the atomic core region corresponds
to the binding energy distribution
\begin{equation}
p_{WCA}(u_{C})=\frac{H(u_{C}-\tilde{u}_{C})(1+\tilde{x}_{C})^{1/2}}{4\epsilon_{LJ}x(1+x)^{3/2}}\label{eq:PC(u)}
\end{equation}
where $H(\cdot)$ is Heavyside's step function, $x=\sqrt{u_{C}/\epsilon_{LJ}}$,
and $\tilde{x}_{C}=\sqrt{\tilde{u}_{C}/\epsilon_{LJ}}$. Here $\tilde{u}_{C}>0$
is an adjustable energy parameter that defines the level set of the
boundary of the core of receptor atoms. It is defined as the repulsive
energy above which the energy of the collision follows the probability
density (\ref{eq:PC(u)}). 

We show in the Appendix that, under reasonable assumptions, the form
(\ref{eq:PC(u)}) of the probability density for the collisional binding
energy component applies for a receptor composed of many atoms, albeit
perhaps with values of parameters $\epsilon_{LJ}$ and $\tilde{u}_{C}$
not obviously related to the assumed Lennard-Jones form of the repulsive
potential. 

We now turn to the generalization of the binding energy probability
distribution for a polyatomic ligand. In this case, because it is
now the sum over both multiple ligand atoms and receptor atoms, the
distribution of the background component $u_{B}$ is expected to obey
the central limit theorem to an even greater extent and, consequently,
it is expected to continue to be well described by the Gaussian form
(\ref{eq:PB(u)}), where now the parameters $\bar{u}_{B}$ and $\sigma_{B}$
refer to average binding energy and corresponding standard deviation
for the whole ligand rather than a single atom. 

Even though the total collisional energy is the sum of the collisional
energies of each ligand atom, the central limit theorem is not applicable
because the mean and variances of each contribution, described by
probability density (\ref{eq:PC(u)}), are undefined. We can assume
however that the collisional energy is dominated by the largest repulsive
interaction among all of the ligand atoms $u_{C}\simeq\max{}_{i=1,N}[u_{C}(i)]$,
where $u_{C}(i)$ is the collisional energy of ligand atom $i$. The
probability density of the maximum, $x_{{\rm max}}$, of a set of
$N$ independent random variables $x_{i}$ distributed according to
the probability density $f(x)$ is given by the expression\cite{GumbelBook}

\begin{equation}
p(x_{{\rm max}})=N\left[F(x_{{\rm max}})\right]^{N-1}f(x_{{\rm max}})\label{eq:pmax(u)}
\end{equation}
where $F(x)$ is the integrated form of $f(x)$, that is the cumulative
distribution corresponding to $f(x)$. In general, the positions of
the $N$ atoms of the ligand are not independent so Eq.~(\ref{eq:pmax(u)})
is an approximation. It is expected however that this form, with an
effective number of statistically independent number of atoms groups,
$n_{l}$, is of general applicability. If the ligand is small and
rigid it will behave as a single atom. On the other extreme, a large
and flexible ligand can be thought of being composed of groups of
atoms with nearly uncorrelated position. 

Combining Eqs.~(\ref{eq:pcu_ratio-1}), (\ref{eq:PC(u)-2-1}), and
(\ref{eq:pmax(u)}) yields
\begin{equation}
p_{WCA}(u_{C})=\frac{1}{\mathcal{N}}\left[1-\frac{(1+x_{C})^{1/2}}{(1+x)^{1/2}}\right]^{n_{l}-1}\frac{H(u_{C}-\tilde{u}_{C})}{4\epsilon_{LJ}}\frac{(1+x_{C})^{1/2}}{x(1+x)^{3/2}}\label{eq:pwcaf(u)}
\end{equation}
where $\mathcal{N}$ is the normalization factor and the other symbols
have the same meaning as in Eq.~(\ref{eq:PC(u)}). 

The probability density (\ref{eq:pwcaf(u)}) is the probability density
of the collisional energy conditional on there being a collision.
That is of being at least one atomic clash defined as $u>\tilde{u}_{C}$.
Denoting by $p_{{\rm core}}$ the probability that one such collision
occurs when the ligand is within the binding site volume, and by $\tilde{p}_{{\rm core}}=1-p_{{\rm core}}$
its complement, we obtain the following expression for the probability
density of the collisional energy: 
\begin{equation}
p_{C}(u_{C})=p_{{\rm core}}p_{WCA}(u_{C})+\tilde{p}_{{\rm core}}\delta(0)\label{eq:PC(u)-1}
\end{equation}
where $p_{WCA}(u_{C})$ is given by Eq.~(\ref{eq:pwcaf(u)}) and
the $\delta$-function expresses the fact that outside the core region
the collisional energy is zero.

Finally, assuming that the background and collisional contributions
are statistically independent, the probability density of the total
binding energy $u=u_{B}+u_{C}$ is given by the convolution of the
respective probability densities:

\begin{equation}
p_{0}(u)=p_{0}(u_{C}+u_{B})=\int_{-\infty}^{+\infty}p_{C}(u')p_{B}(u-u')du',\label{eq:convolution}
\end{equation}
where $p_{B}(u)$ is a Gaussian distribution (\ref{eq:PB(u)}) and
the collisional density given by (\ref{eq:PC(u)-1}). Substituting
these definitions in Eq.~(\ref{eq:convolution}) we obtain:
\begin{equation}
p_{0}(u)=\tilde{p}_{{\rm core}}g(u;\bar{u}_{B},\sigma_{B})+p_{{\rm core}}\int_{\tilde{u}_{C}}^{+\infty}p_{WCA}(u')g(u-u';\bar{u}_{B},\sigma_{B})du'\:,\label{eq:p0(u)conv2}
\end{equation}
where $g(u;\bar{u}_{B},\sigma_{B})$ is the Gaussian distribution
of mean $\bar{u}_{B}$ and standard deviation $\sigma_{B}$ {[}see
Eq.~(\ref{eq:PB(u)}){]}. While the integral in Eq.~(\ref{eq:p0(u)conv2})
is not available in analytical form, it is amenable to numerical computation
by for example Gauss-Hermite quadrature. Fig.~\ref{fig:p0u-1} shows
$p_{0}(u)$ for a particular choice of the parameters $\bar{u}_{B}$,
$\sigma_{B}$, $\epsilon_{LJ}$, $\tilde{u}_{c}$, and $\tilde{p}_{{\rm core}}$.
Also shown in this figure is $p_{1}(u)\propto e^{-u}p_{0}(u)$ {[}see
Eq.~(\ref{eq:plu}){]}. These distributions indeed reflect the behavior
of binding energy distributions obtained from actual molecular simulations
(see Results).

\subsection{Model for the free energy profile}

Since the Laplace transform of a convolution of two functions is the
product of their Laplace transforms, from Eq.~(\ref{eq:Klambdadef-prob-linear})
and Eqs.~(\ref{eq:PC(u)}) and (\ref{eq:PB(u)}), we have
\begin{equation}
K(\lambda)=K_{C}(\lambda)K_{B}(\lambda),\label{eq:klambda1}
\end{equation}
where
\begin{equation}
K_{C}(\lambda)=\int_{-\infty}^{+\infty}p_{C}(u)e^{-\lambda u}du=\tilde{p}_{{\rm core}}+p_{{\rm core}}K_{WCA}(\lambda)\label{eq:Kc-1}
\end{equation}
where $K_{WCA}(\lambda)$ is the two-sided Laplace transform of $p_{WCA}(u)$.
From Eq.~(\ref{eq:pwcaf(u)}):
\begin{equation}
K{}_{WCA}(\lambda)=\int_{\tilde{u}_{C}}^{+\infty}p_{WCA}(u)e^{-\lambda u}du\label{eq:Kc-2}
\end{equation}
Finally, the two-sided Laplace transform of $p_{B}(u)=g(u;\bar{u}_{B},\sigma_{B})$
(a Gaussian) is:

\begin{equation}
K_{B}(\lambda)=\int_{-\infty}^{+\infty}p_{B}(u)e^{-\lambda u}=e^{-\sigma_{B}^{2}\lambda(\lambda-2\bar{u}_{B}/\sigma_{B}^{2})}\label{eq:kblambda}
\end{equation}

\begin{figure}
\begin{centering}
\includegraphics[scale=0.5]{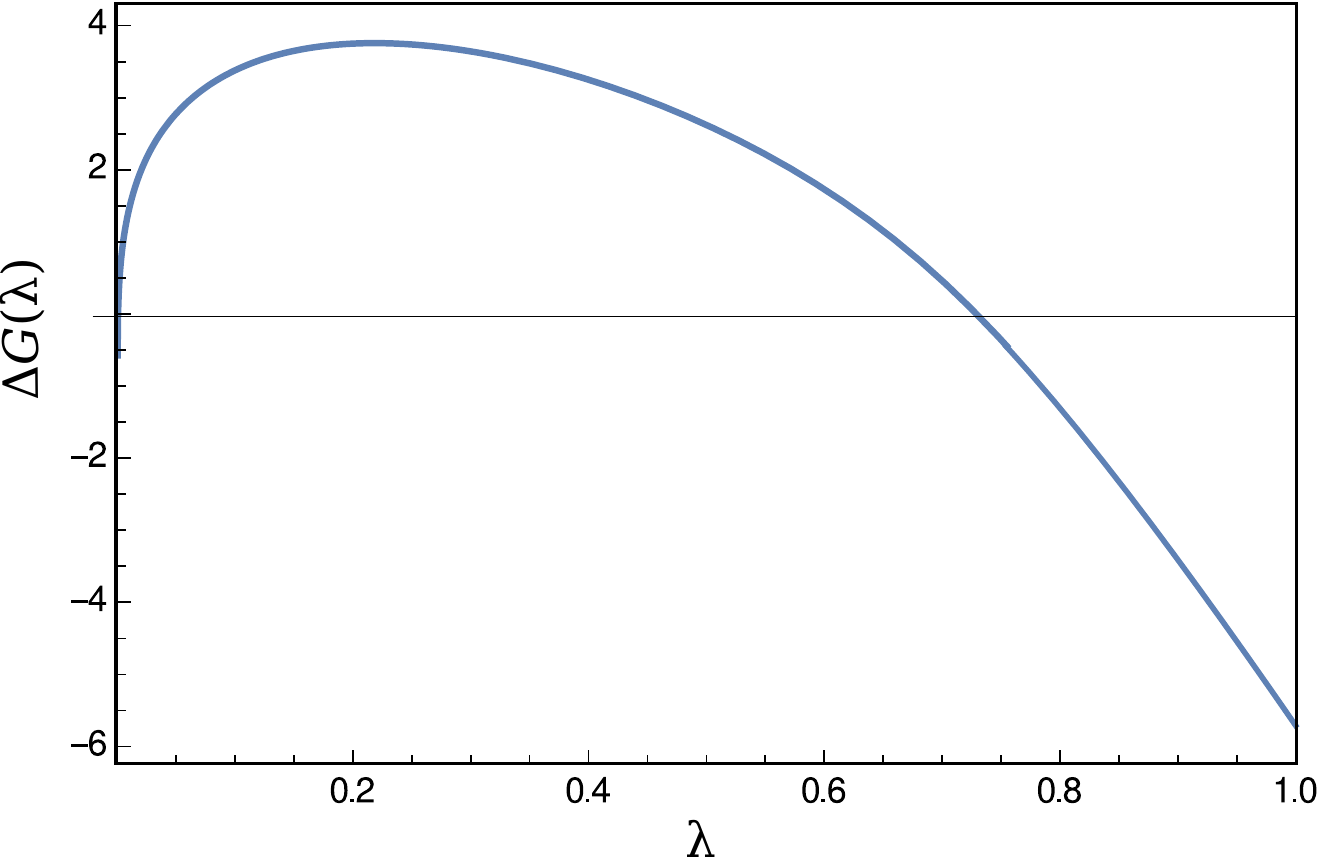}
\par\end{centering}

\caption{\label{fig:klambda1}The binding free energy $\Delta G(\lambda)=-\ln K(\lambda)$
from Eqs.~(\ref{eq:klambda1})--(\ref{eq:kblambda}) as a function
of $\lambda$ for $\bar{u}_{B}=-10$, $\sigma_{B}=3$ $\epsilon_{LJ}=1$,
$\tilde{u}_{c}=10$, $n_{l}=2$ and $\tilde{p}_{{\rm core}}=10^{-6}$.
Energy values are expressed in units of $k_{B}T$.}
 
\end{figure}
An illustrative binding free energy profile, $\Delta G(\lambda)=-\ln K(\lambda)$,
obtained from Eqs.~(\ref{eq:klambda1}), (\ref{eq:Kc-1}), (\ref{eq:Kc-2})
and (\ref{eq:kblambda}) for some choice of parameter values is shown
in Fig.~\ref{fig:klambda1}. Free energy profiles from simulations
indeed follow have the shape illustrated in Fig.~\ref{fig:klambda1}
(see Results). Note that in this model $\Delta G(\lambda)$ is given
by the sum of the free energies corresponding to the collisional and
background processes:
\begin{equation}
\Delta G(\lambda)=-\ln K_{C}(\lambda)-\ln K_{B}(\lambda)=\Delta G_{C}(\lambda)+\Delta G_{B}(\lambda).\label{eq:analytic-free-energy-profile}
\end{equation}

\subsection{Mixture model of background component\label{sub:Mixture-Model}}

The analytical model described so far predicts Gaussian-distributed
binding energies at $\lambda\simeq1$, where the collisional contribution
is negligible. We have encountered, however, systems displaying bimodal
binding energy distributions in this regime (see for example Fig.~\ref{fig:plambdas-etc-alanine}).
These occurrences are interpreted as the system undergoing a conformational
transition from a high-entropy/high-energy state to a low-entropy/low-energy
state as $\lambda$ is increased. We found that these systems can
be described well by a mixture model of the background binding energy
component described by the weighted sum of two Gaussian distributions:
\begin{equation}
p_{B}(u_{B})=P_{a}g(u_{B};\bar{u}_{a},\sigma_{a})+P_{b}g(u_{B};\bar{u}_{b},\sigma_{b})\,,\label{eq:PB(u)-mixture}
\end{equation}
where $P_{a}$ and $P_{b}$ ($P_{a}+P_{b}=1$) are the probabilities
of occurrence of conformational states $a$ and $b$ at $\lambda=0$,
respectively, and $(\bar{u}_{a},\sigma_{a})$ and $(\bar{u}_{a},\sigma_{a})$
are the corresponding average and standard deviation parameters of
their background components at $\lambda=0$. 

To formulate the full model of $p_{0}(u)$ for this case, Eq.~(\ref{eq:PB(u)-mixture})
replaces the single Gaussian $g(u;\bar{u}_{B},\sigma_{B})$ in Eq.~(\ref{eq:p0(u)conv2}).
The remainder of the analytical theory is unchanged. Note that this
model can be expanded to an arbitrary number of states and that it
reduces to the single-state model {[}Eq.~(\ref{eq:PB(u)}){]} when
only one state is present (that is $P_{a}=1$, for example).

\subsection{Model parameterization}

The analytical model of binding defined by Eq.~(\ref{eq:p0(u)conv2})
with Eqs.~(\ref{eq:pwcaf(u)}) and (\ref{eq:PB(u)}) depends on six
parameters: $\bar{u}_{B}$, the average background binding energy
in the coupled state, $\sigma_{B}$, the standard deviation of the
background binding energy in the decoupled state, $\epsilon_{LJ}$,
the effective Lennard-Jones $\epsilon$ parameter of the repulsive
potential within the atomic core, $\tilde{u}_{c}$, the repulsive
energy above which the collisional binding energy contribution is
dominated by the closest contact, $n_{l}$, the effective number of
statistically independent atom groups of the ligand, and $\tilde{p}_{{\rm core}}$,
the probability that in the uncoupled state the ligand does not overlap
with receptor atoms.

These parameters are obtained for each complex by varying them so
as to fit Eq.~(\ref{eq:p0(u)conv2}) to histograms of binding energies
observed in alchemical molecular simulations at multiple values of
$\lambda$ (see Fig.~\ref{fig:plambdas-etc-3} as an example). Near
$\lambda=1$, where ligand-receptor interactions are established,
atomic collisions are unlikely and the binding energy is mainly determined
by the background component. Thus, histograms obtained from molecular
dynamics trajectories near $\lambda=1$ are most useful in the estimation
of the background binding energy parameters $\bar{u}_{B}$ and $\sigma_{B}$.
An initial first guess for these parameters can be extracted from
the average $\langle u_{B}\rangle_{\lambda=1}$ and standard deviation
$\sqrt{\langle\delta u_{B}^{2}\rangle_{\lambda=1}}$ of the binding
energies at $\lambda=1$, observing that, because the background energy
is assumed to be Gaussian-distributed, its parameters follow linear
response behavior upon variation of $\lambda$:
\begin{equation}
\langle\delta u_{B}^{2}\rangle_{\lambda}=\langle\delta u_{B}^{2}\rangle_{0}=\sigma_{B}^{2}\label{eq:sigma-lambda}
\end{equation}

\begin{equation}
\langle u_{B}\rangle_{\lambda}=\langle u_{B}\rangle_{0}-\lambda\sigma_{B}=\bar{u}_{B}-\lambda\sigma_{B}^{2}\:,\label{eq:u-av-lambda}
\end{equation}
which can be easily derived by applying the potential distribution
theorem {[}Eq.~(\ref{eq:plu}){]} to the Gaussian distribution of
$u_{B}$ at $\lambda$: $g[u_{B};\bar{u}_{B}(\lambda),\sigma_{B}(\lambda)]\propto\exp[-\lambda u]g(u_{B};\bar{u}_{B},\sigma_{B})$.

Conversely, the histograms at small $\lambda$ values are most useful
to estimate the collisional energy parameters $\epsilon_{LJ}$, $\tilde{u}_{c}$,
$n_{l}$, and $\tilde{p}_{{\rm core}}$, once a first guess for the
values of $\bar{u}_{B}$ and $\sigma_{B}$ is available. We observed
(see Results), as it would be expected, a high degree of universality
of the parameters $\epsilon_{LJ}$ and $\tilde{u}_{c}$, which describe
the extent and softness of the repulsive potential within the atomic
cores common to all complexes investigated. We used the $\tilde{p}_{{\rm core}}$
parameter, which regulates the relative magnitude of the two components
in Eq.~(\ref{eq:p0(u)conv2}), to match the shape of histograms at
intermediate values of $\lambda$. Finally, we employed the $n_{l}$
parameter to reproduce the shape of the high energy tail of histograms
at small $\lambda$ values (with larger $n_{l}$ values describing
slower decaying tails). Given the difficulty of binning the unbound
high energy portion of binding energies, this last step was performed
by also matching at the shape of the free energy profile $\Delta G(\lambda)$
at small $\lambda$. 

The mixture model (Section \ref{sub:Mixture-Model}) introduces three
additional parameters of the background energy model (the relative
occupancy of the two states, and one additional set of average and
standard deviation parameters of the background component). These
are best obtained by fitting the bimodal distribution of binding energies
as a function of $\lambda$, exploiting the linear response behavior
of each of the average and standard deviation parameters {[}Eqs.~(\ref{eq:sigma-lambda})
and (\ref{eq:u-av-lambda}){]}, and those of the state probabilities:
\begin{equation}
P_{a}(\lambda)=\frac{P_{a}e^{-(\bar{u}_{a}^{2}-\langle u_{a}\rangle_{\lambda}^{2})/2\sigma_{a}^{2}}}{M(\lambda)}
\end{equation}
\begin{equation}
P_{b}(\lambda)=1-P_{a}(\lambda)
\end{equation}
where
\begin{equation}
M(\lambda)=P_{a}e^{-(\bar{u}_{a}^{2}-\langle u_{a}\rangle_{\lambda}^{2})/2\sigma_{a}^{2}}+P_{b}e^{-(\bar{u}_{b}^{2}-\langle u_{b}\rangle_{\lambda}^{2})/2\sigma_{b}^{2}}\:.
\end{equation}
which can be derived by application of the potential distribution
theorem to the Gaussian mixture distribution (\ref{eq:PB(u)-mixture}).

In future work we will implement an automated mechanism based on maximum
likelihood statistical inference to estimate the parameters of the
model.\cite{lee2012new}

\subsection{Computational details}

The host-guest complexes were prepared as described.\cite{Wickstrom2013,Gallicchio2014octacid,pal2016SAMPL5}
Single-decoupling\cite{Gallicchio2010} Hamiltonian Replica-exchange
Molecular dynamics simulations \cite{gallicchio2015asynchronous}
employed 22 intermediate $\lambda$ steps as follows: $\lambda$ =
$0$, $1\times10^{-6}$, $1\times10^{-5}$, $1\times10^{-4}$, $1\times10^{-3}$,
$0.002$, $0.004$, $0.008$, $0.01$, $0.02$, $0.04$, $0.07$,
$0.1$, $0.17$, $0.25$, $0.35$, $0.5$, $0.6$, $0.7$, $0.8$,
$0.9$, and $1$. The calculation employed the OPLS-AA force field\cite{Jorgensen:Maxwell:96,Kaminski:2001}
and the AGBNP2 implicit solvent model.\cite{Gallicchio2009} The replica-exchange
(AsyncRE) simulations were started from energy-minimized and thermalized
structures from manually docked models. A flat-bottom harmonic restraint
with a tolerance of $5$ Å between the centers of mass of the host
and the guest was applied to define the binding site volume. Each
cycle of a replica lasted for 100 picoseconds with 1 fs time-step.
The average sampling time for a replica was approximately 10 ns. Calculations
were performed on the campus computational grid at Brooklyn College.
The binding energies obtained from all replicas were analyzed using
UWHAM\cite{Tan2012} method and the R-statistical package to compute
the binding free energy profile $\Delta G_{{\rm b}}(\lambda)$.

\section{Results}

\begin{figure}
\begin{centering}
\begin{minipage}[t]{0.3\columnwidth}%
\begin{center}
$\beta$-cyclodextrin/\\
cyclohexanol
\par\end{center}

\begin{center}
\includegraphics[scale=0.17]{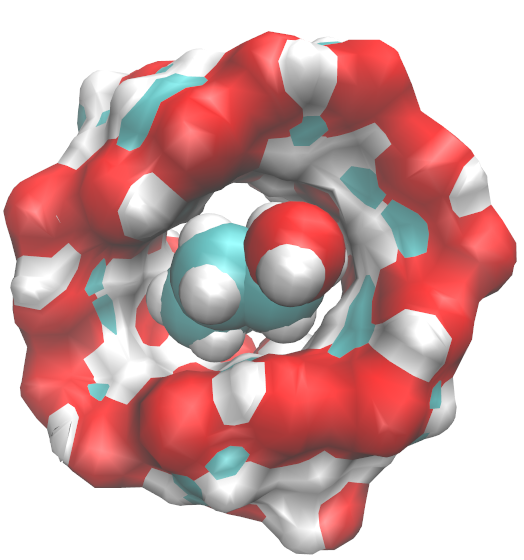}
\par\end{center}%
\end{minipage}%
\begin{minipage}[t]{0.3\columnwidth}%
\begin{center}
$\beta$-cyclodextrin/\\
nabumetone
\par\end{center}

\begin{center}
\includegraphics[scale=0.2]{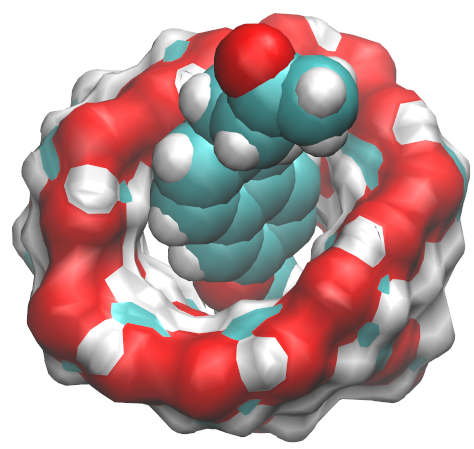}
\par\end{center}%
\end{minipage}%
\begin{minipage}[t]{0.35\columnwidth}%
\begin{center}
octa-acid cavitand/\\
trans-4-methylcyclohexanoate
\par\end{center}

\begin{center}
\includegraphics[scale=0.2]{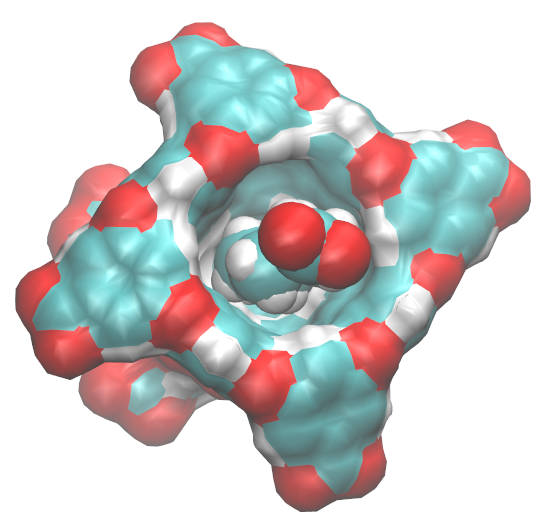}
\par\end{center}%
\end{minipage}
\par\end{centering}

\caption{Molecular representations of three of the four host-guest complexes
studied in this work. The host is shown in surface representation
and the guest is shown using van der Waals atomic spheres. \label{fig:hg1}}

\end{figure}

\begin{figure}
\begin{centering}
$\beta$-cyclodextrin/cyclohexanol
\par\end{centering}

\begin{centering}
\includegraphics{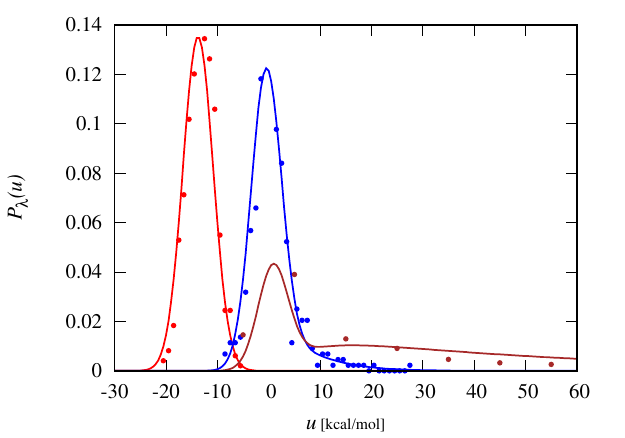}\includegraphics{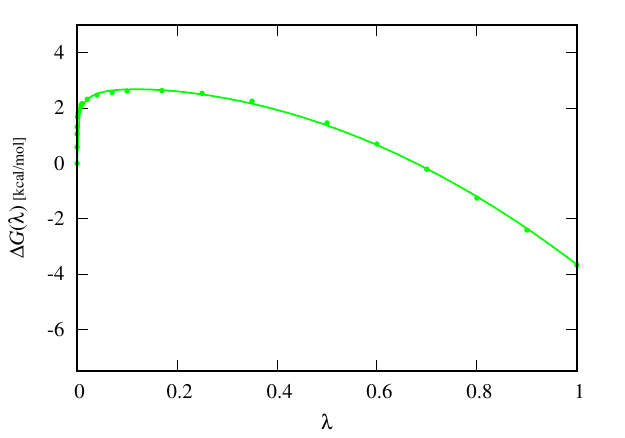}
\par\end{centering}

\begin{centering}
$\beta$-cyclodextrin/nabumetone
\par\end{centering}

\begin{centering}
\includegraphics{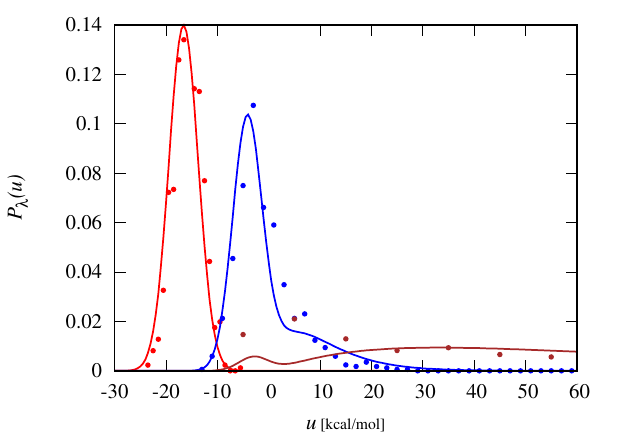}\includegraphics{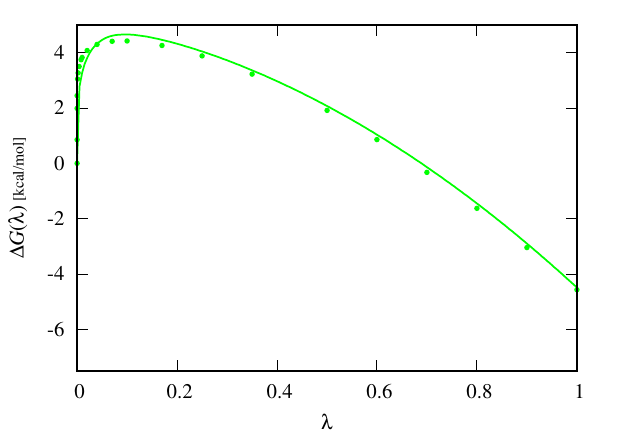}
\par\end{centering}

\begin{centering}
octa-acid/trans-4-methylcyclohexanoate
\par\end{centering}

\begin{centering}
\includegraphics{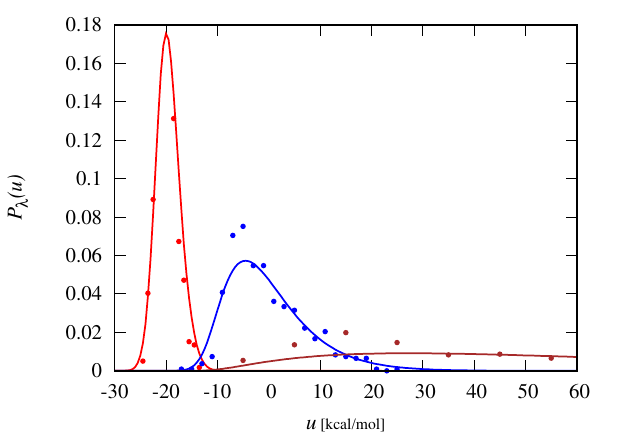}\includegraphics{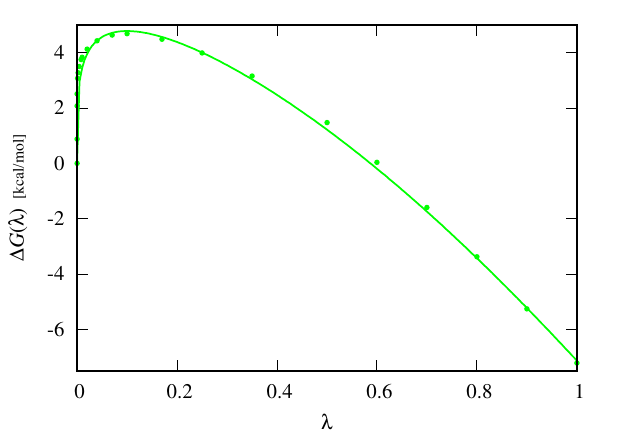}
\par\end{centering}

\caption{Binding energy probability densities $p_{\lambda}(u)$ and binding
free energy profiles for the complexes of cyclohexanol and nabumetone
with $\beta$-cyclodextrin and of trans-4-methylcyclohexanoate with
the octa-acid cavitand. Binding energy probability densities are shown
for (from left to right) for $\lambda=1$ (red), $\lambda=0.1$ (blue),
and $\lambda=0.01$ (brown) with corresponding histogram estimates
from alchemical molecular calculations (filled circles). Analytical
binding free energy profiles (right, green) are compared to UWHAM
numerical estimates from the analysis of alchemical molecular simulation
results.\label{fig:plambdas-etc-3}}
\end{figure}

\begin{figure}
\begin{centering}
$\beta$-cyclodextrin/N-tBOC-L-alanine
\par\end{centering}

\begin{centering}
\begin{minipage}[t]{0.3\columnwidth}%
\begin{center}
state $a$
\par\end{center}

\begin{center}
\includegraphics[scale=0.17]{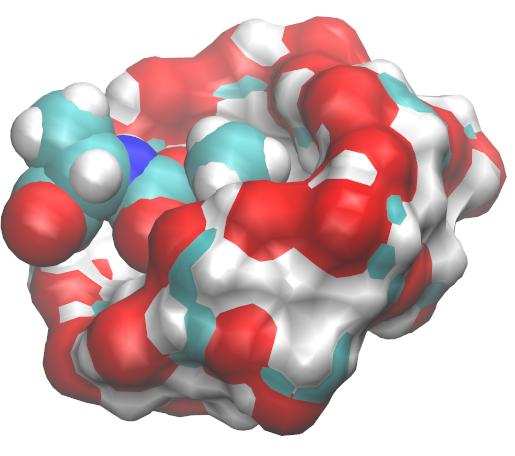}
\par\end{center}%
\end{minipage}%
\begin{minipage}[t]{0.3\columnwidth}%
\begin{center}
state $b$
\par\end{center}

\begin{center}
\includegraphics[scale=0.18]{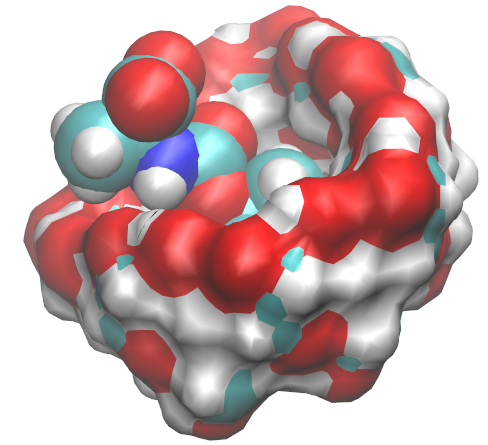}
\par\end{center}%
\end{minipage}\caption{Molecular representations of the two conformations of the$\beta$-cyclodextrin/N-tBOC-L-alanine
representative of the conformational states $a$ and $b$ discussed
in the text. State $b$ (right), in which the carboxylate group is
oriented toward the solvent, is characterized by a more favorable
binding energy than state $a$. However, the configurations of the
guest and host in state $a$ are many times more likely than state
$b$ in absence of guest/host interactions. The complex undergoes
a transition from state $a$ to state $b$ as $\lambda$ increases.
\label{fig:hg2}}

\par\end{centering}

\end{figure}

\begin{figure}
\begin{centering}
$\beta$-cyclodextrin/N-tBOC-L-alanine
\par\end{centering}

\begin{centering}
\includegraphics{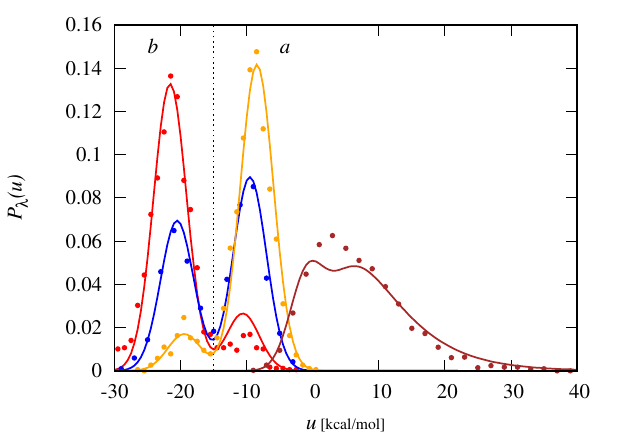}\includegraphics{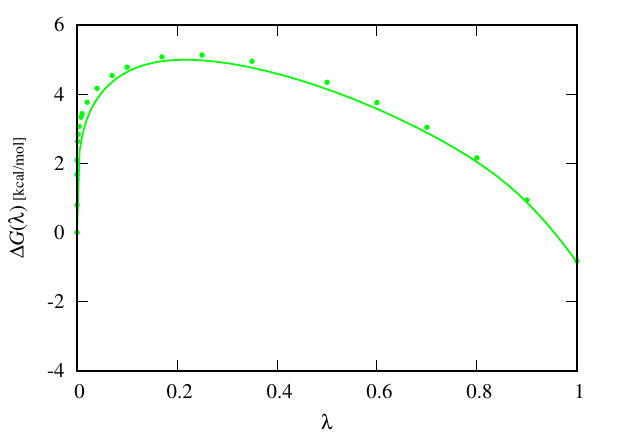}
\par\end{centering}

\caption{Binding energy probability densities $p_{\lambda}(u)$ and binding
free energy profiles for the complex of N-tBOC-L-alanine with $\beta$-cyclodextrin.
Binding energy probability densities are shown for (from left to right)
for $\lambda=1$ (red), $\lambda=0.9$ (blue), $\lambda=0.8$ (orange),
and $\lambda=0.1$ (brown) with corresponding histogram estimates
from alchemical molecular calculations (filled circles). A transition
from a high binding energy state $a$ to a low binding energy state
$b$ (see Fig.~\ref{fig:hg2}) occurs at $\lambda\simeq0.9$. The
vertical dotted line separates the probability density peaks characteristic
of the two states. Analytical binding free energy profiles (right,
green) are compared to UWHAM numerical estimates from the analysis
of alchemical molecular simulation results.\label{fig:plambdas-etc-alanine}}
\end{figure}

We tested the analytical model of binding presented above on four
host-guest complexes: cyclohexanol, nabumetone, and N-tBOC-L-alanine
binding to $\beta$-cyclodextrin\cite{Wickstrom2013} and trans-4-methylcyclohexanoate
binding to the octa-acid cavitand host\cite{pal2016SAMPL5} (Figs.~\ref{fig:hg1}
and \ref{fig:hg2}). The results for the complexes with cyclohexanol,
nabumetone, and trans-4-methylcyclohexanoate are shown in Fig.~\ref{fig:plambdas-etc-3}
and Table \ref{tab:parameters-3}. The results for the complex of
N-tBOC-L-alanine and $\beta$-cyclodextrin, which undergoes a $\lambda$-dependent
conformational transition, are presented in Fig.~\ref{fig:plambdas-etc-alanine}
and Table \ref{tab:parameters-alanine}.

\begin{table}
\begin{centering}
\caption{Model parameters for the complexes of cyclohexanol and nabumetone
with $\beta$-cyclodextrin and of trans-4-methylcyclohexanoate with
the octa-acid cavitand.\label{tab:parameters-3}}
\begin{tabular}{lccccccc}
 & $\Delta G_{b}^{\circ}$$^{a}$ & $\bar{u}_{B}$$^{a}$ & $\sigma_{B}$$^{a}$ & $\tilde{p}_{{\rm core}}$  & $\tilde{u}_{c}$$^{a}$ & $\epsilon_{LJ}$$^{a}$  & $n_{l}$\tabularnewline
\hline 
{\small{}cyclohexanol} & $-3.0$ & $1.0$ & $2.95$ & $1.0\times10^{-2}$ & $0.5$ & $20$ & $3.5$\tabularnewline
{\small{}nabumetone} & $-3.9$ & $-2.8$ & $2.85$ & $1.5\times10^{-4}$ & $0.5$ & $20$ & $5.5$\tabularnewline
{\small{}trans-4-methylcyclohexanoate} & $-6.5$ & $-14.0$ & $1.93$ & $3.0\times10^{-8}$ & $0.5$ & $20$ & $6.0$\tabularnewline
\hline 
\end{tabular}
\par\end{centering}

$^{a}${\footnotesize{}In kcal/mol}
\end{table}

The analytic model fits very well the binding energy distributions
and free energy profiles from numerical calculations (Fig.~\ref{fig:hg1}).
The model correctly interpolates the Gaussian behavior of the binding
energy distributions at $\lambda\simeq1$ and the diffuse and asymmetric
aspects of the distributions at $\lambda\simeq0$. As shown in Fig.~\ref{fig:hg1},
the distributions at intermediate $\lambda$ values present characteristics
of both limits and are also correctly described by the model.

Free energy profiles (Fig.~\ref{fig:hg1}, right column) are also
closely described by the analytic model. For large values of $\lambda$
($\lambda>0.3$, approximately), the free energy profiles vary quadratically
with $\lambda$, consistent with linear response behavior. The quadratic
regime is preceded by a highly non-linear variation of the free energy
near $\lambda=0$. The analytic model correctly captures the singularity
of the first derivative of the free energy profile at $\lambda=0^{+}$.\cite{simonson1993free}
The maximum of the free energy corresponds to the value of $\lambda$
at which the average binding energy is zero. In general, as it can
be shown from Eqs.~(\ref{eq:TI}) and (\ref{eq:dudl=00003Du}), the
first derivative of the free energy profile is in fact proportional
to the average binding energy. The singularity of the first derivative
at $\lambda=0^{+}$ is, thus, consistent with the undefined first
moment of the $p_{0}(u)$ probability density. As the data in Fig.~\ref{fig:hg1}
illustrates, the analytic model successfully interpolates between
the linear response regime at $\lambda\simeq1$ and the collisional
regime at $\lambda\simeq0$.

The model parameters obtained by fitting the analytic predictions
to the numerical results for three of the four host-guest complexes
are listed in Table \ref{tab:parameters-3}. The values of the standard
binding free energies (2nd column) differ from the corresponding free
energy profiles (Fig.~\ref{fig:plambdas-etc-3}) at $\lambda=1$
by the standard state concentration-dependent term (see Methods).
The stronger binding affinities of nabumetone and trans-4-methylcyclohexanoate
to their respective hosts relative to cyclohexanol is driven by stronger
interaction energies as reflected by the $\bar{u}_{B}$ parameter.
The average binding energies at the bound state $\lambda=1$ match
closely the linear response predictions from Eq.~(\ref{eq:u-av-lambda}):
$\langle u\rangle_{1}=-13.7$, $-16.6$, and $-20.3$ kcal/mol, from
Eq.~(\ref{eq:u-av-lambda}) and fitted $\bar{u}_{B}$, $\sigma_{B}$
parameters (Table \ref{tab:parameters-3}), for cyclohexanol, nabumetone
and trans-4-methylcyclohexanoate, respectively, compared to the direct
numerical estimates $\langle u\rangle_{1}=-13.2$, $-15.7$, and $-20.0$
kcal/mol, from direct numerical averaging of the binding energies
from the $\lambda=1$ simulation replicas. 

The trend toward stronger interaction energies is partially offset
by the progressively smaller probabilities of fitting the guest into
the host without causing atomic clashes, as illustrated by the $\tilde{p}_{{\rm core}}$
parameter (Table \ref{tab:parameters-3}, 5th column). For example,
the estimates indicate that it is almost 6 orders of magnitude more
difficult to fit trans-4-methylcyclohexanoate into the octa-acid cavitand
that it is to fit cyclohexanol into $\beta$-cyclodextrin. This presumably
reflects the fact that the $\beta$-cyclodextrin interior is larger
than that of the octa-acid cavitand, which, in addition, is open only
on one end. The variations of $\tilde{p}_{{\rm core}}$ could also
represent the probabilities of occurrence of binding-competent conformations
of guest and host. 

As expected, a common set of values of the $\tilde{u}_{c}$ and $\epsilon_{LJ}$
parameters, corresponding loosely to the magnitude and softness of
the core inter-atomic repulsion potential, describes all of the complexes
investigated. The magnitude of the fitted $\epsilon_{LJ}$ parameter
($\epsilon_{LJ}=20$ kcal/mol) is significantly larger than typical
Lennard-Jones $\epsilon$ force field parameters. This confirms the
expectation that these parameters should be interpreted to represent
the shape and intensity of the repulsive potential exercised by groups
of atoms, rather than by individual atoms. 

Finally, in Table \ref{tab:parameters-3} we report the fitted values
of the the $n_{l}$ parameter (8th column) which represents the number
of statistically independent number of atom groups of the guests.
Indeed, $n_{l}$ values roughly scale as the size of the guest. For
example nabumetone binding to $\beta$-cyclodextrin corresponds to
$n_{l}=5.5$, whereas the smaller cyclohexanol has $n_{l}=3.5$. Despite
the smaller size, the $n_{l}$ value of trans-4-methylcyclohexanoate
binding to the octa-acid cavitand is similar to that of nabumetone,
possibly reflecting an influence of the nature of the receptor on
the value of this parameter.

\begin{table}
\begin{centering}
\caption{Model parameters for the complexes of N-tBOC-L-alanine with $\beta$-cyclodextrin.\label{tab:parameters-alanine}}
\begin{tabular}{lccccccc}
N-tBOC-L-alanine & $P_{{\rm state}}$ & $\bar{u}_{B}$$^{a}$ & $\sigma_{B}$$^{a}$ & $\tilde{p}_{{\rm core}}$  & $\tilde{u}_{c}$$^{a}$ & $\epsilon_{LJ}$$^{a}$  & $n_{l}$\tabularnewline
\hline 
{\small{}state $a$} & $\sim1.0$ & $0$ & $2.5$ & \multirow{2}{*}{$8.9\times10^{-5}$} & \multirow{2}{*}{$0.5$} & \multirow{2}{*}{$20$} & \multirow{2}{*}{$4.5$}\tabularnewline
{\small{}state $b$} & $4.0\times10^{-8}$ & $-11$ & $2.5$ &  &  &  & \tabularnewline
\hline 
\end{tabular}
\par\end{centering}

$^{a}${\footnotesize{}In kcal/mol}
\end{table}

The complex of N-tBOC-L-alanine with $\beta$-cyclodextrin (Fig.~\ref{fig:hg2})
undergoes a transition along the alchemical path from one conformational
state (state $a$ in Fig.~\ref{fig:hg2}), in which the carboxylate
terminus is hydrogen-bonded to one of the hydroxyl groups of the wide
rim of the host, to another conformational state (state $b$ in Fig.~\ref{fig:hg2}),
in which the carboxylate group is rotated toward the solvent, and
the body of the aminoacid, including the tert-butyl moiety, is seated
deeper in the host interior than in state $a$. 

The transition is particular evident in the distributions of binding
energy values as a function of $\lambda$ (Fig.~\ref{fig:plambdas-etc-alanine}).
For the other complexes studied (Fig.~\ref{fig:plambdas-etc-3})
the peaks of the binding energy distributions linearly shift toward
more negative values as $\lambda$ is increased. In contrast, the
peak of the distribution for N-tBOC-L-alanine, starting at $\lambda\simeq0.8$,
instead of shifting, it develops a low energy peak as $\lambda$ is
increased. Between $\lambda=0.8$ and $\lambda=1$ the distribution
is bimodal, with a high energy mode (corresponding to state $a$)
centered near $u=-9$ kcal/mol and the low energy mode (corresponding
to state $b$) near $u=-22$ kcal/mol. Starting at $\lambda=0.8$,
where state $a$ is predominant, population abruptly shifts to state
$b$, which becomes the predominant state at $\lambda=1$. At $\lambda=0.9$
the two states have almost the same population. This behavior is the
hallmark of a pseudo first-order phase equilibrium,\cite{Kim2010,binder2015generalized}
in which two phases, characterized by compensating differences in
average energy and entropy, cohexist within the same free energy range. 

Indeed, the mixture model (Section \ref{sub:Mixture-Model} and Table
\ref{tab:parameters-alanine}) captures the tread off between interaction
energy (the $\bar{u}_{B}$ parameter) and probability of occurrence
(the $P_{{\rm state}}$ parameter). State $a$ has a much higher probability
of occurrence in the uncoupled state than state $b$. However in state
$b$ the guest interacts more favorably with the host than in state
$a$ by about $-11$ kcal/mol (Table \ref{tab:parameters-alanine},
3rd column). For small $\lambda$ values, the binding energy component
of the alchemical potential energy function {[}Eq.~(\ref{eq:U-lambda-linear}){]}
is small and the complex tends to visit exclusively state $a$ given
its overwhelmingly large probability. However, as $\lambda$ is increased,
the weight of the binding energy term increases and state $b$ becomes
competitive with state $a$.

The conformational transition is also apparent in the abrupt change
of slope of the binding free energy profile near $\lambda=0.9$ (Fig.~\ref{fig:plambdas-etc-alanine}).
As mentioned, the slope of the binding free energy profile corresponds
to the average binding energy as a function of $\lambda$. Correspondingly,
at $\lambda\simeq0.9$, the system transitions to a state of lower
binding energy thereby causing the change in slope. Note that, while
the transition appears slight in the binding free energy profile,
the shift in slope causes a significant decrease (by about 1 kcal/mol)
of the binding free energy. Furthermore, the shift in slope of the
binding free energy profile and the bimodal character of the binding
energy distributions can not be described without invoking the mixture
model.

\section{Discussion}

The results obtained as part of this work clearly indicate that it
is feasible to represent alchemical binding free energy profiles by
parameterized analytic functions. The model offers a rationalization
for the shape of the free energy profile and of the binding energy
distributions. The critical feature of the model is the ability to
bridge the two limiting behaviors of the free energy profile, the
region near $\lambda\simeq0$ determined by atomic clashes and the
region near $\lambda\simeq1$ characterized by linear response. The
main conceptual advance that enabled this versatility of the model
is the description of the binding energy in the uncoupled state of
the complex as the sum of two interaction energy components with radically
distinct statistical signatures. The first, termed ``collisional''
interaction energy, describes atomic clashes dominated by nearest
neighbor pairs and follows the statistics of the maximum of a set
of random variables. The second, that we termed ``background'' interaction
energy, describes the sum of many weak and favorable interatomic interactions
and follows the central limit theorem. The two statistical components,
assumed statistically independent, are then combined using standard
convolution to obtain the distribution of the total binding energy
and, by means of a Laplace transformation, the binding free energy
profile.

The general strategy of describing free energy changes along a thermodynamic
path by means of probability models applied to the ``decoupled''
end point has a long history in the treatment of solvation phenomena
in condensed phases. Examples are scaled particle theory, particle
insertion models, and information/fluctuation theories.\cite{reiss1959statistical,Widom1982,Pratt1992,Hummer:Pratt:96,Huang:Chandler:2002}
Early work in this area by Pratt \& Chandler,\cite{Pratt:Chandler:77}
introduced the connection between the solubility of hard sphere particles\cite{stamatopoulou1998cavity}
and the probability of formation of suitable cavities in the neat
solvent, a prediction that was confirmed by Pangali, Rao, and Berne\cite{Pangali:Rao:Berne:79b}
and subsequent computer simulation work.\cite{Wallqvist:Berne:94,Berne:96,Gallicchio:Kubo:Levy:2000}
Both Pohorille and Pratt\cite{Pohorille1990a} and Hummer et al.,\cite{Hummer:Pratt:96}
elaborated on the concept of, $p_{0}(r)$, the probability that a
cavity of size $r$ occurs in a neat liquid, which was first introduced
in scaled particle theory\cite{reiss1959statistical,Pierotti:76}
to model the probability of occurrences of cavities based on the moments
of the number of solvent molecules that occupy the solute volume in
neat water. 

The same essential concepts have been used here to formulate a model
connecting the free energy of inserting a ligand molecule into a receptor
binding site to probability distributions collected in the decoupled
state. The main difference between the solvation process, seen as
solute insertion, and binding, seen as ligand insertion, is that,
unlike a homogeneous solution, the distribution of receptor atoms
is not homogeneous. In particular, there are regions in the receptor
binding site where a ligand can fit without requiring conformational
reorganization. Conversely, there are interior regions of the receptor
from where the ligand is effectively excluded. The model we formulated
takes into account these complex geometric and energetic effects in
terms of effective physical parameters which are set so as to reproduce
the results of alchemical molecular simulations. The close agreement
obtained here between model predictions and molecular simulations
of a set of relatively simple but yet chemically-relevant host-guest
complexes, is evidence that the model is sound and deserving of further
investigation and development.

The physical parameters returned by the model can be useful in the
characterization of molecular complexes. For instance, the $\bar{u}_{B}$
and $\sigma_{B}$ parameters measure the strength of favorable electrostatic
and dispersion receptor-ligand interactions as a function of $\lambda$
{[}Eq.~(\ref{eq:u-av-lambda}){]}. In particular, the $\sigma_{B}$
parameter measures the linear response of the complex to the establishment
of favorable interactions. A larger $\sigma_{B}$ can be an indication,
for example, of a larger polarizability of the receptor and can be
interpreted in terms of local dielectric constant.\cite{archontis2001dielectric,simonson2002gaussian,simonson2008dielectric,nymeyer2008method}
On the other hand, the $\tilde{p}_{{\rm core}}$ parameter, which
is the probability that ligand and receptor do not overlap, is a measure
of the size of the binding cavity, if present, relative to the size
of the ligand, or, alternatively, the likelihood of the formation
of a suitable binding cavity that can fit the ligand. Similarly, the
$n_{l}$ parameter can be taken as a measure of ligand size and ligand
flexibility. As discussed, the mixture model parameters indicate the
presence of multiple conformational states of the complex and of their
average interaction energies and relative probabilities. Taken together,
these parameters, when tabulated over a series of systems, can be
useful to characterize and categorize receptor-ligand complexes and,
when correlated with binding affinities, can inform receptor and ligand
design.

Future work will also assess the potential usefulness of the analytic
model toward the improvement of alchemical simulation protocols. A
possible application of the model is as a framework to analyze and
measure free energy changes near the decoupled state without the need
for extrapolation\cite{deng2011elucidating} or soft-core alchemical
potentials.\cite{Tan2012,Buelens2012} As analyzed by Simonson\cite{simonson1993free}
and reproduced by our model, the first derivative of the free energy
profile has a singularity at $\lambda=0$. This singularity causes
problems for numerical free energy estimators,\cite{Shirts2005,Shirts2008a}
which are usually addressed by the adoption of non-linear soft-core
alchemical potentials.\cite{Pohorille2010,shirts2013introduction}
These difficulties can also be addressed by replacing the numerical
estimation of free energies near the singularity with the estimation
of the parameters (which are free of singularities) of the analytic
free energy function (\ref{eq:analytic-free-energy-profile}). The
analytic model can also be potentially useful to evaluate alchemical
thermodynamic lengths to optimize the $\lambda$ schedule\cite{schon1996thermodynamic_distance,shenfeld2009minimizing}
of alchemical transformations.

The model, as currently expressed, is limited to single-decoupling
linear alchemical transformations.\cite{Gallicchio2011adv,di2015unique}
Single-decoupling requires pre-averaging to the solvent degrees of
freedom by means of a solvent potential of mean force treatment\cite{Roux:Simonson:99}
implemented here using the AGBNP2\cite{Gallicchio2009} implicit solvent
model. The requirement of linearity of the alchemical transformation
with respect to the charging parameter $\lambda$ is introduced so
as to deploy potential distribution theorem identities\cite{PDTbook:2006}
relating binding energy distributions at different values of $\lambda$.
Future work will attempt to extend the model to non-linear coupling
schemes and to explicit solvation models. Binding free energy calculations
with explicit solvation are typically conducted according to the double-decoupling
scheme,\cite{Gilson:Given:Bush:McCammon:97} which is based on the
difference of the free energies of coupling the ligand to the hydrated
receptor and the free energy of solvation. Hence, it is conceivable
that an analogous analytic model can be developed for double-decoupling
alchemical calculations by considering each free energy leg separately.

\section{Conclusion}

We have presented a parameterized analytical model describing the
free energy profile of linear single-decoupling alchemical binding
free energy calculations. The parameters of the model, which are physically
motivated, are obtained by fitting model predictions to numerical
simulations. The validity of the model has been assessed on a set
of host-guest complexes. The model faithfully reproduces the binding
free energy profiles and the probability densities of the perturbation
energy as a function of the alchemical progress parameter $\lambda$.
The model offers a rationalization for the characteristic shape of
the free energy profiles. The parameters obtained from the model are
potentially useful descriptors of the association equilibrium of molecular
complexes.

\newpage{}


\newpage{}

\section{Appendix}

\subsection{Derivation of Eq.~(\ref{eq:PC(u)})}

\begin{figure}
\begin{centering}
\includegraphics[scale=0.5]{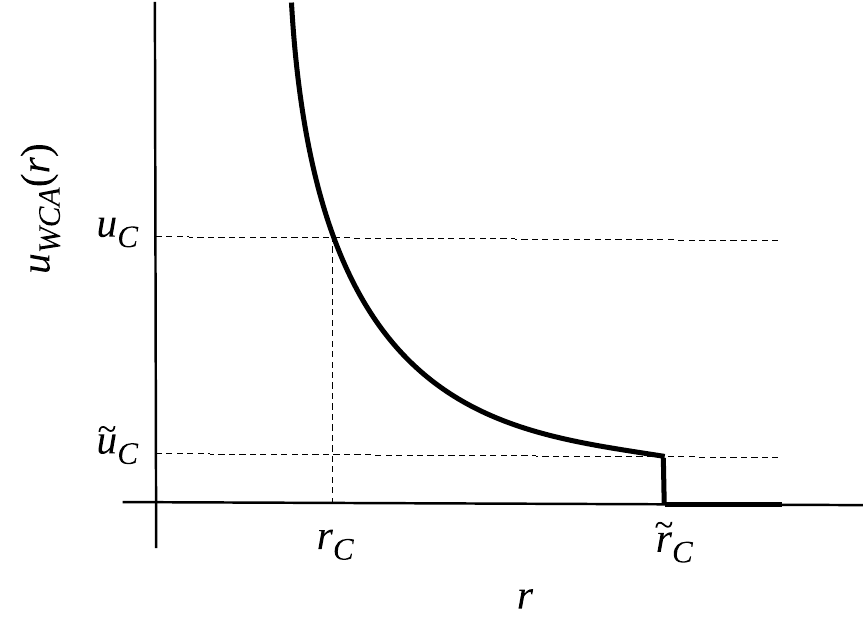}\caption{Representation of the repulsive WCA component of the Lennard-Jones
potential {[}Eq.~(\ref{eq:WCApot}){]} used in the derivation of
Eq.~(\ref{eq:PC(u)}). Here $\tilde{r}_{C}$ represents the radius
of the spherical core region around a receptor atom and $\tilde{u}_{C}$
the corresponding repulsion potential energy. Similarly, $r_{C}$
is a generic distance between the ligand atom and the receptor atom
within the core and $u_{C}$ is the corresponding potential energy.\label{fig:wca_pot}}

\par\end{centering}

\end{figure}

Consider two particles interacting by the pair potential (\ref{eq:WCApot})
in which one particle (representing the receptor) is fixed at the
origin and the other (representing the ligand) is uniformly distributed
in a sphere of radius $r_{C}$ centered at the origin (Fig.~\ref{fig:wca_pot}).
Here we assume that $r_{c}<r_{0}=2^{1/6}\sigma_{LJ}$ (the distance
beyond which the Lennard-Jones WCA potential is zero). We will derive
the probability density $p_{C}(u_{C})$ of the interaction energy
$u_{{\rm WCA}}(r)$, where $r$ is the distance between the two particles,
by differentiating the cumulative probability function $P_{C}(u_{C})$
defined as the probability that, given that the ligand particle is
uniformly distributed in the sphere, the interaction energy $u_{{\rm WCA}}(r)$
is greater than the given value $u_{C}$. The value of the WCA potential
at $r_{c}$ is denoted by $\tilde{u}_{c}$; $\tilde{u}_{c}$ is therefore
the smallest allowed interaction energy.

The probability that the pair interaction energy is smaller than $u_{C}$
is given by:
\begin{equation}
P_{WCA}(u_{C})=H(u_{C}-\tilde{u}_{c})\frac{V_{C}-V(u_{C})}{V_{C}}\label{eq:pcu_ratio-1}
\end{equation}
where the Heaviside function imposes the requirement that $u_{C}$
be larger than the minimum values, $V_{C}$ is the volume of the sphere
of radius $r_{c}$ and $V(u_{C})$ is the volume of the sphere of
radius $r(u_{C})$, where $r(u_{C})$ is inter-particle distance at
which the LJ WCA potential has value $u_{C}$. From Eq.~(\ref{eq:WCApot})
we have 
\begin{equation}
r(u_{C})=\frac{r_{0}}{(1+x_{C})^{1/6}};\quad u\ge0\label{eq:ru-2}
\end{equation}
where $r_{0}=2^{1/6}\sigma_{LJ}$ is the minimum of the Lennard-Jones
pair potential and
\begin{equation}
x_{C}=\sqrt{u_{c}/\epsilon_{LJ}}\label{eq:x-def}
\end{equation}
Inserting Eq.~(\ref{eq:ru-2}) into Eq.~(\ref{eq:pcu_ratio-1})
and differentiating with respect to $u_{C}$ yields Eq.~(\ref{eq:PC(u)}),
which expresses a normalized distribution as it can be verified by
direct integration using the fact that 
\begin{equation}
\int\frac{dx}{(1+x)^{3/2}}=\frac{2}{(1+x)^{1/2}}\label{eq:ru-1-1}
\end{equation}

Now consider a receptor composed of $M$ atoms interacting with a
monoatomic ligand with the WCA repulsive potential (\ref{eq:WCApot}).
The cumulative probability is given by the expression $P_{WCA}(u_{C})=H(u_{C}-\tilde{u}_{c})(1-V(u_{C})/V_{C})$,
as in Eq.~(\ref{eq:pcu_ratio-1}), where now $V(u_{C})$ is the volume
of the region of the receptor where the WCA potential is larger than
$u_{C}$ and, similarly, $V_{C}\ge V(u_{C})$ is the volume where
the WCA potential is larger than $\tilde{u}_{C}$. We can approximate
$V(u_{C})$ by the van der Waals volume $V[r(u_{C})]$ of a molecule
with $M$ atoms with van der Waals radii $r(u_{C})$, given by Eq.~(\ref{eq:ru-2})
below, corresponding to distance at which the value of WCA repulsive
pair potential is equal to $u_{C}$. Differentiating the cumulative
distribution with respect to $u_{C}$, yields:
\begin{equation}
p_{WCA}(u_{C})=-\frac{1}{V_{C}}\frac{dV(r)}{dr}\frac{dr(u_{C})}{du_{C}}=\frac{H(u_{C}-\tilde{u}_{C})}{12\epsilon_{LJ}}\frac{A(r)r}{V_{C}}\frac{1}{x(1+x)^{3/2}}\label{eq:PC(u)-2}
\end{equation}
where $A(r)$ is the van der Waals surface of the receptor when the
atomic radii are set to $r$, and $r$ and $x$ are both functions
of $u_{C}$ {[}see Eqs.~(\ref{eq:ru-2}) and (\ref{eq:x-def}){]}.

Eq.~(\ref{eq:PC(u)-2}) is interesting because it links the probability
density of the collisional interaction energy to the shape of the
receptor. There are numerical algorithms (some analytical) to obtain
the van der Waals surface area of a molecule. For large $u$, $r(u)$
is small and atomic overlaps between receptor atoms can be ignored.
In this limit $A[r(u)]\simeq M4\pi r(u)^{2}$, and assuming that $V_{C}\simeq M4\pi r(\tilde{u}_{C})^{3}/3$,
we obtain 
\begin{equation}
p_{WCA}(u_{C})=\frac{H(u_{C}-\tilde{u}_{C})}{4\epsilon_{LJ}}\frac{(1+x_{C})^{1/2}}{x(1+x)^{3/2}}\label{eq:PC(u)-2-1}
\end{equation}
which has the same form as the probability density of the collisional
energy for one receptor atom.
\end{document}